\documentstyle[12pt,epsfig]{article}
\newcommand{\be}{\begin{equation}}  
\newcommand{\ee}{\end{equation}}
\newcommand{\ba}{\begin{eqnarray}}  
\newcommand{\ea}{\end{eqnarray}}
\newcommand{\m}{_{\mu}}
\newcommand{\amu}{A_{\mu}}
\newcommand{\id}{{\sl 1} \!\!{\sl I}}
\def\I {{\rm 1} \hspace{-1.1mm} {\rm I} \hspace{0.5mm}}

\title{\bf Lattice gauge-fixing\\
           for generic covariant gauges}
\author{
{L.~Giusti}\\
{\small\it Scuola Normale Superiore, P.zza dei Cavalieri N.7 - 56100
Pisa Italy}\\[-0.2cm]
{\small\it INFN Sezione di Pisa, 56100 Pisa Italy}\\[-0.2cm]
{\small\it Dip. di Fisica, Universit\`a degli Studi di Roma 
"La Sapienza"}\\[-0.2cm]
{\small\it P.le A.~Moro 2 - 00185 Rome  Italy.}  \\ 
 \\
}
 
\begin{document}

%\begin{flushright}
%CERN-TH.?????\\
%ROME prep. 94/1022 \\
%??normale
%\end{flushright}

\date{}
\maketitle

\thispagestyle{empty} 

\abstract{We propose a method which allows the generalization of the Landau 
lattice
gauge-fixing procedure to generic covariant gauges. We report preliminary
numerical results showing how the procedure works for $SU(2)$ and $SU(3)$. We
also report numerical results showing that the contribution of finite
lattice-spacing effects and/or spurious copies are relevant in the lattice
gauge-fixing procedure.}
\pagestyle{empty}\clearpage
\newpage
\pagestyle{plain} \setcounter{page}{1}
\section{Introduction}
Recently, lattice QCD Monte Carlo simulations requiring a gauge-fixing  
have become relevant; the gauge-fixing is essential to 
study gauge dependent quantities like, for example, the propagators of the  
fundamental fields entering the continuum QCD lagrangian and to use smearing 
techniques \cite{bacilieri}.\\ 
The study of gluon and quark propagators, among other things, allows a better  
understanding of the infrared behaviour of the theory and the confinement  
mechanism. Moreover, quark and gluon matrix elements can be used to obtain  
renormalization conditions as proposed in \cite{maiani,guido1}.\\ 
In the last few years, numerical studies of lattice propagators have been 
performed by  
several groups: the aim of the authors \cite{mandula2}-\cite{bernard1} was to
study the mechanism  
through which the gluon may become massive at long distances whereas 
more recent  
attempts studied its behaviour as a function of momentum 
\cite{nico1}-\cite{bernard2}. 
Analogous studies of the quark propagator exist\cite{bernard1}.\\
The existence of lattice Landau and Coulomb gauge-fixing ambiguities has been
verified \cite{claudio1}-\cite{paciello1}. Studying the characteristics of these
ambiguities and their influences on gauge fixed quantities is interesting at
least for two reasons: the existence of these ambiguities could be the analogous
of the Gribov problem in the continuum formulation of non abelian gauge
theories \cite{gribov} and the gauge-fixing is essential for the analytical
study of the continuum limit of lattice gauge theories.\\ 
In practice, there are some cases in which it is convenient to use a gauge 
dependent  
procedure to compute gauge invariant quantities. For example, smeared 
fermionic  
interpolating operators are being used in lattice QCD spectroscopy and  
phenomenology. The smearing operators are gauge dependent and therefore the 
gauge must be fixed before they are calculated.\\ 
Up to now, the only covariant gauge for which it is known the algorithm to fix
it on the lattice is the Landau Gauge \cite{wilson1}-\cite{davies1}.
This algorithm uses the original idea of
Gribov  \cite{gribov} in the continuum, restricting the domain of integration 
of the
partition function in the region where the functional 
$F(\Omega)\equiv ||A^\Omega||^2$ reaches
an extreme. Also algorithms for non
covariant gauges are discussed in literature (see for example \cite{livio}).\\
In this paper we propose a procedure which allows to generalize the Landau
gauge-fixing on the lattice for a generic covariant gauge. 
This gauge-fixing procedure could verify the gauge independence of some 
results obtained in
literature \cite{mandula2}-\cite{bernard2} and could allow to discriminate 
between gauge artifacts and true physical
properties of the fundamental fields entering the QCD lagrangian.     

\section{Covariant gauges quantization}\label{par2}
In this section we will briefly review the general formalism to quantize a non
abelian gauge theory using covariant gauge conditions \cite{books}.\\
Neglecting the Gribov problem,  
let us assume that we can find a gauge section in the space of gauge fields
which intersects once and only once all gauge orbits. The Landau gauge is
readily extended to an auxiliary gauge-fixing condition of the form
\be\label{dinamic1}
\partial\m\amu^\Omega(x)=\Lambda(x)\; 
\ee
where $\Lambda(x)$ belongs to the Lie algebra of the
group. Since gauge-invariant quantities should not be sensitive to changes of
gauge condition, it is possible to average over $\Lambda(x)$ with a
gaussian weight 
\be\label{o2}
Z(J_{\cal O})=\int\delta\Lambda e^{-\frac{1}{2 \alpha}\int d^4x Tr(\Lambda^2)}
\int \delta\amu \delta\eta \delta\bar{\eta} e^{-S(A)-
S_{ghost}(\eta,\bar{\eta},\amu)+\int J_{\cal O} {\cal O}}
\delta(\partial\m\amu-\Lambda)
\ee
obtaining the standard formula
\be\label{o3}
Z(J_{\cal O})=\int \delta\amu \delta \eta \delta \bar{\eta} e^{-S(A)-
S_{ghost}(\eta,\bar{\eta},\amu)+\int J_{\cal O} {\cal O}}
e^{-\frac{1}{2 \alpha}\int d^4x(\partial_{\mu}A_{\mu})( 
\partial_{\mu}A_{\mu})}\; .\nonumber
\ee
where ${\cal O}$ is a gauge-invariant operator.\\
The choice $\alpha=1$ is referred to as Feynman gauge instead for $\alpha=0$
the Landau gauge is recovered.\\
In the next sections we will show how to implement such a formulation for a non
perturbative numerical simulation to compute the mean value of a
gauge-dependent observable on the lattice using the expression (\ref{o2}) for
$Z(J_{\cal O})$. 

\section{The functional for covariant gauges}\label{confix}
The functional proposed by Gribov, directly in the continuum, in order 
to fix the Landau Gauge is
\be
F(\Omega)\equiv ||A^\Omega||^2=
\int\mbox{\rm Tr}\left(A^\Omega_{\mu}(x)A^\Omega_{\mu}(x)\right)d^4x
\ee
where
\be\label{contgt}
A^\Omega_{\mu}(x)=\Omega(x) A_{\mu}(x)\Omega^{\dagger}(x)-\frac{i}{g}\Omega(x)
\partial_{\mu}\Omega^{\dagger}(x)\;.  
\ee
and
\be\label{gaugetr}
\Omega(x)=e^{iw(x)}
\ee
is a group matrix, $w(x)$ and $A_\mu(x)$ belonging to the Lie algebra of 
the group.\\
In appendix A we show that 
\be\label{deltaf}
\frac{\delta F(\Omega)}{\delta w^b}=-\frac{2}{g}(\partial_{\mu}
A^\Omega_{\mu})^a\Phi^{ab}(w) 
\ee
with
\be\label{cov10}
\Phi^{ab}(w) \equiv  \left[\frac{e^{\gamma} - \I }{\gamma}\right]^{ab}\;
\;\;
\gamma^{ab} \equiv  f^{abc} w^c\;\; ;
\ee
Equation (\ref{deltaf}) shows that $F(\Omega)$ is stationary when
$\partial_{\mu}A^\Omega_{\mu}=0$.\\
In order to fix the gauge discussed in the section \ref{par2}
we should be able to find a functional $H(\Omega)$ stationary when
\be
\partial_{\mu}A^{\Omega}_{\mu}(x)=\Lambda(x)\; , 
\ee
with $\Lambda(x)$ having a gaussian distribution.
The most naive way to define $H(\Omega)$ would be to find a functional 
$h(\Omega)$ such that 
\be\label{latcov1}
\frac{\delta h}{\delta
w^b(x)}=\frac{2}{g}\left(\Lambda^a(x)\Phi^{ab}(w(x))\right)
\ee
so that 
\be
\frac{\delta (h + F)}{\delta w^b(x)}=-\frac{2}{g}
\left(\partial_{\mu} A_{\mu}^\Omega(x)-\Lambda(x))^a\Phi^{ab}(w(x)\right)\;\; .
\ee 
However we will now show that this is not possible; in fact for a non abelian 
gauge theory does not exist a functional satisfying (\ref{latcov1}).
A {necessary} condition for the existence of such a 
functional would be
\be
\frac{\delta^2 H(\Omega)}{\delta w^c(x)\delta w^b(y)}= 
\frac{\delta^2 H(\Omega)}{\delta w^b(y)\delta w^c(x)}
\ee
which implies the integrability condition
\be\label{cov3}
\frac{\delta}{\delta w^c(x)}(\Lambda^a(y)\Phi^{ab}(w))=
\frac{\delta}{\delta w^b(y)}(\Lambda^a(x)\Phi^{ac}(w))\; .
\ee
Expanding $\Phi^{ab}(w(x))$ in power of $w(x)$, equation (\ref{cov3}) 
should be satisfied order by order in $w(x)$. From equation (\ref{cov10}) we 
have
\be
\Phi^{ab}(w)\simeq \delta^{ab} +\frac{\gamma}{2}^{ab}
=\delta^{ab}+f^{abc}\frac{w^c}{2}\; ,
\ee
Equation (\ref{cov3}) is then in contrast with the antisymmetry of $f^{abc}$.\\ 
The functional we propose in order to fix the gauge discussed in section 
\ref{par2} is:
\be\label{cov11}
H(\Omega)=\int d^4x\mbox{\rm Tr}\left[(\partial_{\mu}A^\Omega_{\mu}-\Lambda)  
(\partial_{\nu}A^\Omega_{\nu}-\Lambda)\right]\; .  
\ee
In fact
\be
\frac{\delta H(\Omega)}{\delta w^a}=2\Phi^{ab}(w)\left[
D_{\nu}\partial_{\nu}(\partial_{\mu}A^\Omega_{\mu}-\Lambda)\right]^b
\ee
which shows that $H(\Omega)$ is stationary when 
\be\label{ngf}
\partial_{\mu}A^\Omega_{\mu}(x)-\Lambda(x)=0\;\; . 
\ee
The problem is to show that no spurious stationary points of
(\ref{cov11}) exist. In fact, for example, apparently
\be\label{theo1}
\partial_{\mu}A^\Omega_{\mu}(x)-\Lambda(x)=\mbox{\it cost}\; . 
\ee 
also satisfies
\be\label{bestia1}
\frac{\delta H(\Omega)}{\delta w^a}=0
\ee
and other non trivial solutions of the equation (\ref{bestia1}) could exist 
as well. In our case we can exclude the possibility
of constant zero modes. In fact integrating the 
equation (\ref{theo1}) we obtain
\[
\int\partial_{\mu}A^\Omega_{\mu}(x)d^4x-\int\Lambda(x)d^4x=\mbox{\it cost}\int
d^4x\; .
\]
Through the use of periodic boundary conditions we have
\be\label{nuova1}
- \int\Lambda(x) d^4 x =\mbox{\it cost}\int d^4 x\; .
\ee 
Equation (\ref{nuova1}) assures that $cost=0$ when $\Lambda(x)$ has a
gaussian distribution. This result assures 
that the functional $H(\Omega)$ does not have spurious stationary points
satisfying equation (\ref{theo1}).\\ 
It is interesting to note that for $\Lambda(x)=0$ the functional
$H(\Omega)$ has absolute minima for any $\Omega$ satisfying 
$\partial_{\mu}A^\Omega_{\mu}(x)=0$. This is not the case for the Gribov
functional.

\section{Covariant gauges on the Lattice}\label{quarto}
The gauge variables of a compact lattice gauge theory are the links $U_\mu(x)$
and they are elements of the gauge group.\\
In this section we outline a procedure to compute numerically the
mean value of a gauge dependent operator $\cal{O}$ on the lattice using a
generic covariant gauge quantization. The mean value of a gauge dependent
operator on the lattice is defined as :
\be\label{omedio}
\langle{\cal O}\rangle=\int\delta\Lambda e^{-\frac{1}{2 \alpha}\int d^4x 
Tr(\Lambda^2)}
\int dU \Delta_f(U)\delta(\Delta-\Lambda) 
e^{-S_W(U)} {\cal O}(U)\; .
\ee
where $S_W(U)$ is the Wilson lattice gauge invariant action, 
$\Delta_f(U)$ is the Faddeev-Popov determinant and   
\be
\Delta(x)=\frac{1}{2iag}\sum_{\mu=1}^{4}
[U_\mu(x)-U_\mu^{\dagger}(x)]_{traceless}-
[U_\mu(x-\mu)-U_\mu^{\dagger}(x-\mu)]_{traceless}\;\; .
\ee
The computation of the integral (\ref{omedio}) can be
schematized as follows:
\begin{itemize}
\item  A gauge configuration
$\{U\}$ with periodic boundary conditions according to the gauge invariant 
weight $e^{-S_W(U)}$ is generated;
\item For each $\{U\}$ configuration random matrices
$\Lambda(x)$ belonging to the group algebra are extracted according to 
the weight $e^{-\frac{1}{2\alpha}Tr\Lambda^2(x)}$;
\item Given $\Lambda(x)$, a numerical algorithm extremizes a 
discretization of the functional
$H(\Omega)$. This defines the lattice gauge-fixing condition 
\end{itemize}
\be\label{divefun1}
\Delta(x)-\Lambda(x)=0\;\;\; \forall x\; ;
\ee
\begin{itemize}
\item the mean value of the lattice gauge dependent operator is then defined as
\end{itemize}
\be
\langle{\cal O}\rangle^{Latt}=\frac{1}{N}\sum_{\{conf\}}{\cal O}(U_i)\;\; .
\ee
Since $\Lambda(x)$
has a gaussian distribution for all $x$, if the number of lattice 
sites is sufficiently large
\[
\frac{1}{V}\sum_{x}\Lambda(x)=\bar{\Lambda}=0\; .
\]

\section{Lattice gauge-fixing for covariant gauges}
A possible discretization of $H(\Omega)$ is 
\be\label{lattf}
H_L(\Omega)= \frac{1}{V}\mbox{Tr}\sum_x\left[\Delta^\Omega(x)-
\Lambda(x)\right]^2
\ee
where
\be
\Delta^\Omega(x)=\frac{1}{2iag}\sum_{\mu=1}^{4}
[U_\mu^\Omega(x)-U_\mu^{\Omega \dagger}(x)]_{traceless}-
[U_\mu^\Omega(x-\mu)-U_\mu^{\Omega \dagger}(x-\mu)]_{traceless}\; ,
\ee
$\Lambda(x)$ are matrices which belong to the group algebra, $V$ is the
lattice volume and links transform under a gauge 
transformation as  
\[
U_\mu^\Omega(x)=\Omega(x)U_\mu(x)\Omega^{\dagger}(x+\mu)\;\; .
\]                                                          
$H_L(\Omega)$ is the simplest discretization of $H(\Omega)$. 
For a generic transformation close to $\Omega^0$ we define
\[
\frac{\delta H}{\delta\Omega}\Big|_{\Omega^0}\equiv
\frac{\delta H}{\delta\epsilon}\;\;\;\; ,\;\;\;\; \Omega(x)= \Omega^0(x)(\I - i
\epsilon(x))
\] 
and we have
\begin{eqnarray}\label{lattder}
\frac{\delta H_L}{\delta\epsilon^a}=-\frac{1}{Vga}\sum_\mu\mbox{Tr}
\left[T^a\right.\left(
(\Delta^{\Omega^0}(x)-\Lambda(x))(U^{\Omega^0 \dagger}_{\mu}(x)+\right.
\nonumber\\
U^{\Omega^0}_{\mu}(x)+U^{\Omega^0 \dagger}_{\mu}(x-\mu)+U^{\Omega^0}_{\mu}
(x-\mu)) + \nonumber\\
 -  (\Delta^{\Omega^0}(x-\mu)-\Lambda(x-\mu)) 
(U^{\Omega^0 \dagger}_{\mu}(x-\mu)+U^{\Omega^0}_{\mu}(x-\mu))+\\
\left.(\Delta^{\Omega^0}(x+\mu)-\Lambda(x+\mu)) 
(U^{\Omega^0 \dagger}_{\mu}(x)+U^{\Omega^0}_{\mu}(x))\right)
\left.\right]\;\;\nonumber
\end{eqnarray}
where $T^a$ are the group generators.\\ 
We have to find a gauge transformation $\Omega$ such that $\{U_\mu^\Omega\}$ 
satisfies
\be\label{ddir}
\frac{\delta H_L}{\delta\epsilon^a(x)}=0\;\; \forall \; x\; .
\ee
In order to solve numerically equation (\ref{ddir}) we use an iterative 
algorithm for the minimization of $H_L(\Omega)$. If such an algorithm converges 
(this is 
not guaranteed and must be checked in practice), its fixed 
points will be configurations satisfying the condition (\ref{ddir}).\\
In order to study the convergence of the algorithm, two lattice quantities 
can be monitored
\begin{eqnarray}
H_L & = & \frac{1}{V}\mbox{Tr}\sum_x\left[\Delta(x)-
\Lambda(x)\right]^2 \nonumber\\
\vartheta_H & \equiv & V \sum_{x,a} 
\frac{\delta H_L}{\delta\epsilon^a}
\frac{\delta H_L}{\delta\epsilon^a}\; .
\nonumber
\end{eqnarray}
The function $H_L(\Omega)$ is defined on a compact set. If the numerical algorithm
of minimization converges, $\vartheta_H$ must go to a value of the
order of the precision required for the minimization whatever starting configuration 
we use. The procedure works if also $H_L$ goes to a value of the order of the
precision required for the minimization for those configurations where the 
algorithm converges.
In the next section
we will show numerical results suggesting that $H_L(\Omega)$ goes to zero when 
it is minimized for the configurations considered.\\ 
We remark that the same discretized definition of gauge field 
$(1/2iag)[U_\mu(x)-U_\mu^{\dagger}(x)]_{traceless}$ is used in the definition 
of $H_L(\Omega)$ and of $\Delta^{\Omega}(x)$.\\
On the contrary in the Landau lattice gauge-fixing procedure, usually adopted in the
literature, different definitions of the lattice gauge potential are used .
In section \ref{exappb} we report numerical results showing
interesting phenomena arising if different lattice gauge potential
definitions are used in the lattice Landau gauge-fixing procedure.  
   
\section{Numerical Simulations}\label{numeric1}
In this section we report preliminary results of some numerical simulations. In
this exploratory study we considered $SU(2)$ and $SU(3)$ non thermalized
configurations with links defined as
\[
U_\mu(x)=e^{ibB_\mu(x+\frac{\mu}{2})}
\]
with 
\[
B_\mu(x+\frac{\mu}{2}) = \sum_a B_\mu^a(x+\frac{\mu}{2})T^a
\]
where $B_\mu^a(x+\frac{\mu}{2})$ are smooth functions of $x$ (typically
sin$(x)$ and cos$(x)$), $T^a$ are group generators and "$b$" is a
parameter which determines the distance of $U_\mu(x)$ from identity. We
generated, on a volume $V=4^4$, nine configurations for each group, three for 
each value $b=0.1$, $b=0.3$, $b=0.7$. Since the function $H_L(\Omega)$ is 
defined on a 
compact set, the procedure works whatever
configuration we start from, except for numerical convergence problems. 
A complete study of this numerical 
algorithm for thermalized Monte Carlo configurations will be presented in a 
future paper.\\
The gauge-fixing algorithm implements an iterative minimization for
$H_L(\Omega)$ which updates link matrices via $SU(2)$ subgroups, as
proposed in ref.\cite{cabibbo1}. We monitored the quantities $H_L$ 
and $\vartheta_H$ after every
lattice sweep of the gauge-fixing algorithm. We mention that
after each gauge-fixing sweep we orthogonalized the link variables, to 
make sure that they have not been driven off the gauge group.\\
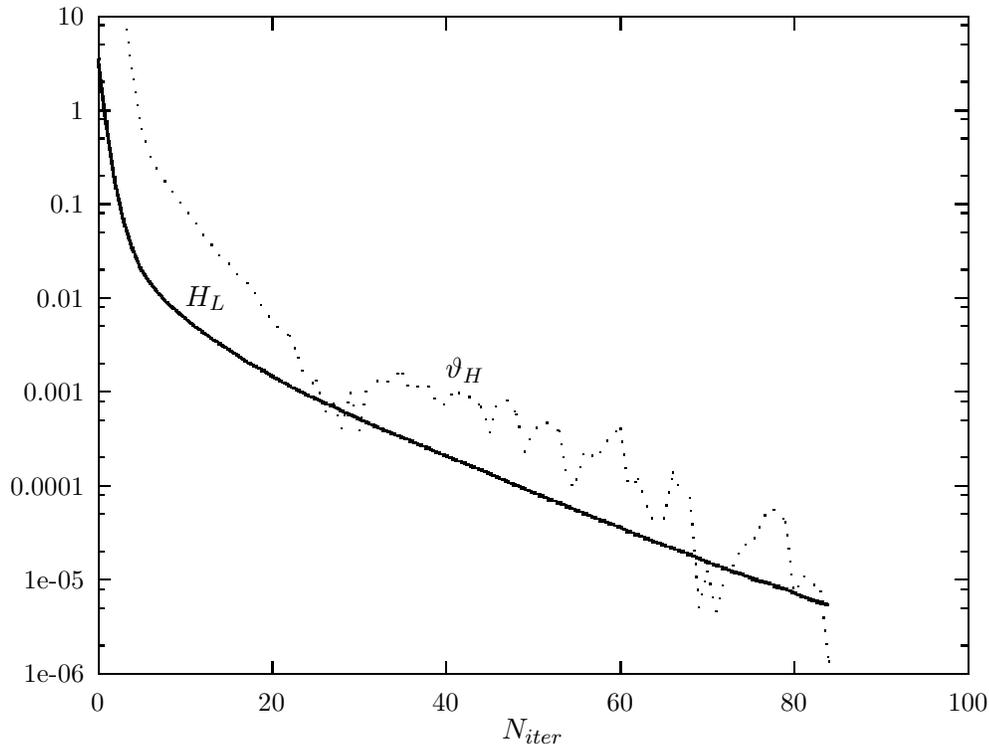
\begin{figure}[htb]   
\begin{center}
% GNUPLOT: LaTeX picture
\setlength{\unitlength}{0.240900pt}
\ifx\plotpoint\undefined\newsavebox{\plotpoint}\fi
\sbox{\plotpoint}{\rule[-0.200pt]{0.400pt}{0.400pt}}%
\begin{picture}(1650,1169)(0,0)
\font\gnuplot=cmr10 at 10pt
\gnuplot
\sbox{\plotpoint}{\rule[-0.200pt]{0.400pt}{0.400pt}}%
\put(220.0,113.0){\rule[-0.200pt]{0.400pt}{248.850pt}}
\put(220.0,113.0){\rule[-0.200pt]{4.818pt}{0.400pt}}
\put(198,113){\makebox(0,0)[r]{1e-06}}
\put(1566.0,113.0){\rule[-0.200pt]{4.818pt}{0.400pt}}
\put(220.0,157.0){\rule[-0.200pt]{2.409pt}{0.400pt}}
\put(1576.0,157.0){\rule[-0.200pt]{2.409pt}{0.400pt}}
\put(220.0,216.0){\rule[-0.200pt]{2.409pt}{0.400pt}}
\put(1576.0,216.0){\rule[-0.200pt]{2.409pt}{0.400pt}}
\put(220.0,246.0){\rule[-0.200pt]{2.409pt}{0.400pt}}
\put(1576.0,246.0){\rule[-0.200pt]{2.409pt}{0.400pt}}
\put(220.0,261.0){\rule[-0.200pt]{4.818pt}{0.400pt}}
\put(198,261){\makebox(0,0)[r]{1e-05}}
\put(1566.0,261.0){\rule[-0.200pt]{4.818pt}{0.400pt}}
\put(220.0,305.0){\rule[-0.200pt]{2.409pt}{0.400pt}}
\put(1576.0,305.0){\rule[-0.200pt]{2.409pt}{0.400pt}}
\put(220.0,364.0){\rule[-0.200pt]{2.409pt}{0.400pt}}
\put(1576.0,364.0){\rule[-0.200pt]{2.409pt}{0.400pt}}
\put(220.0,394.0){\rule[-0.200pt]{2.409pt}{0.400pt}}
\put(1576.0,394.0){\rule[-0.200pt]{2.409pt}{0.400pt}}
\put(220.0,408.0){\rule[-0.200pt]{4.818pt}{0.400pt}}
\put(198,408){\makebox(0,0)[r]{0.0001}}
\put(1566.0,408.0){\rule[-0.200pt]{4.818pt}{0.400pt}}
\put(220.0,453.0){\rule[-0.200pt]{2.409pt}{0.400pt}}
\put(1576.0,453.0){\rule[-0.200pt]{2.409pt}{0.400pt}}
\put(220.0,511.0){\rule[-0.200pt]{2.409pt}{0.400pt}}
\put(1576.0,511.0){\rule[-0.200pt]{2.409pt}{0.400pt}}
\put(220.0,541.0){\rule[-0.200pt]{2.409pt}{0.400pt}}
\put(1576.0,541.0){\rule[-0.200pt]{2.409pt}{0.400pt}}
\put(220.0,556.0){\rule[-0.200pt]{4.818pt}{0.400pt}}
\put(198,556){\makebox(0,0)[r]{0.001}}
\put(1566.0,556.0){\rule[-0.200pt]{4.818pt}{0.400pt}}
\put(220.0,600.0){\rule[-0.200pt]{2.409pt}{0.400pt}}
\put(1576.0,600.0){\rule[-0.200pt]{2.409pt}{0.400pt}}
\put(220.0,659.0){\rule[-0.200pt]{2.409pt}{0.400pt}}
\put(1576.0,659.0){\rule[-0.200pt]{2.409pt}{0.400pt}}
\put(220.0,689.0){\rule[-0.200pt]{2.409pt}{0.400pt}}
\put(1576.0,689.0){\rule[-0.200pt]{2.409pt}{0.400pt}}
\put(220.0,703.0){\rule[-0.200pt]{4.818pt}{0.400pt}}
\put(198,703){\makebox(0,0)[r]{0.01}}
\put(1566.0,703.0){\rule[-0.200pt]{4.818pt}{0.400pt}}
\put(220.0,748.0){\rule[-0.200pt]{2.409pt}{0.400pt}}
\put(1576.0,748.0){\rule[-0.200pt]{2.409pt}{0.400pt}}
\put(220.0,806.0){\rule[-0.200pt]{2.409pt}{0.400pt}}
\put(1576.0,806.0){\rule[-0.200pt]{2.409pt}{0.400pt}}
\put(220.0,837.0){\rule[-0.200pt]{2.409pt}{0.400pt}}
\put(1576.0,837.0){\rule[-0.200pt]{2.409pt}{0.400pt}}
\put(220.0,851.0){\rule[-0.200pt]{4.818pt}{0.400pt}}
\put(198,851){\makebox(0,0)[r]{0.1}}
\put(1566.0,851.0){\rule[-0.200pt]{4.818pt}{0.400pt}}
\put(220.0,895.0){\rule[-0.200pt]{2.409pt}{0.400pt}}
\put(1576.0,895.0){\rule[-0.200pt]{2.409pt}{0.400pt}}
\put(220.0,954.0){\rule[-0.200pt]{2.409pt}{0.400pt}}
\put(1576.0,954.0){\rule[-0.200pt]{2.409pt}{0.400pt}}
\put(220.0,984.0){\rule[-0.200pt]{2.409pt}{0.400pt}}
\put(1576.0,984.0){\rule[-0.200pt]{2.409pt}{0.400pt}}
\put(220.0,998.0){\rule[-0.200pt]{4.818pt}{0.400pt}}
\put(198,998){\makebox(0,0)[r]{1}}
\put(1566.0,998.0){\rule[-0.200pt]{4.818pt}{0.400pt}}
\put(220.0,1043.0){\rule[-0.200pt]{2.409pt}{0.400pt}}
\put(1576.0,1043.0){\rule[-0.200pt]{2.409pt}{0.400pt}}
\put(220.0,1102.0){\rule[-0.200pt]{2.409pt}{0.400pt}}
\put(1576.0,1102.0){\rule[-0.200pt]{2.409pt}{0.400pt}}
\put(220.0,1132.0){\rule[-0.200pt]{2.409pt}{0.400pt}}
\put(1576.0,1132.0){\rule[-0.200pt]{2.409pt}{0.400pt}}
\put(220.0,1146.0){\rule[-0.200pt]{4.818pt}{0.400pt}}
\put(198,1146){\makebox(0,0)[r]{10}}
\put(1566.0,1146.0){\rule[-0.200pt]{4.818pt}{0.400pt}}
\put(220.0,113.0){\rule[-0.200pt]{0.400pt}{4.818pt}}
\put(220,68){\makebox(0,0){0}}
\put(220.0,1126.0){\rule[-0.200pt]{0.400pt}{4.818pt}}
\put(493.0,113.0){\rule[-0.200pt]{0.400pt}{4.818pt}}
\put(493,68){\makebox(0,0){20}}
\put(493.0,1126.0){\rule[-0.200pt]{0.400pt}{4.818pt}}
\put(766.0,113.0){\rule[-0.200pt]{0.400pt}{4.818pt}}
\put(766,68){\makebox(0,0){40}}
\put(766.0,1126.0){\rule[-0.200pt]{0.400pt}{4.818pt}}
\put(1040.0,113.0){\rule[-0.200pt]{0.400pt}{4.818pt}}
\put(1040,68){\makebox(0,0){60}}
\put(1040.0,1126.0){\rule[-0.200pt]{0.400pt}{4.818pt}}
\put(1313.0,113.0){\rule[-0.200pt]{0.400pt}{4.818pt}}
\put(1313,68){\makebox(0,0){80}}
\put(1313.0,1126.0){\rule[-0.200pt]{0.400pt}{4.818pt}}
\put(1586.0,113.0){\rule[-0.200pt]{0.400pt}{4.818pt}}
\put(1586,68){\makebox(0,0){100}}
\put(1586.0,1126.0){\rule[-0.200pt]{0.400pt}{4.818pt}}
\put(220.0,113.0){\rule[-0.200pt]{329.069pt}{0.400pt}}
\put(1586.0,113.0){\rule[-0.200pt]{0.400pt}{248.850pt}}
\put(220.0,1146.0){\rule[-0.200pt]{329.069pt}{0.400pt}}
\put(45,629){\makebox(0,0){$ $}}
\put(903,23){\makebox(0,0){$N_{iter}$}}
\put(357,703){\makebox(0,0)[l]{$H_L$}}
\put(766,590){\makebox(0,0)[l]{$\vartheta_H$}}
\put(220.0,113.0){\rule[-0.200pt]{0.400pt}{248.850pt}}
\sbox{\plotpoint}{\rule[-0.400pt]{0.800pt}{0.800pt}}%
\put(220,1079){\usebox{\plotpoint}}
\multiput(221.41,1051.84)(0.509,-4.192){21}{\rule{0.123pt}{6.543pt}}
\multiput(218.34,1065.42)(14.000,-97.420){2}{\rule{0.800pt}{3.271pt}}
\multiput(235.41,945.46)(0.509,-3.470){19}{\rule{0.123pt}{5.431pt}}
\multiput(232.34,956.73)(13.000,-73.728){2}{\rule{0.800pt}{2.715pt}}
\multiput(248.41,867.46)(0.509,-2.323){21}{\rule{0.123pt}{3.743pt}}
\multiput(245.34,875.23)(14.000,-54.232){2}{\rule{0.800pt}{1.871pt}}
\multiput(262.41,809.97)(0.509,-1.598){21}{\rule{0.123pt}{2.657pt}}
\multiput(259.34,815.48)(14.000,-37.485){2}{\rule{0.800pt}{1.329pt}}
\multiput(276.41,769.51)(0.509,-1.195){19}{\rule{0.123pt}{2.046pt}}
\multiput(273.34,773.75)(13.000,-25.753){2}{\rule{0.800pt}{1.023pt}}
\multiput(289.41,742.19)(0.509,-0.759){21}{\rule{0.123pt}{1.400pt}}
\multiput(286.34,745.09)(14.000,-18.094){2}{\rule{0.800pt}{0.700pt}}
\multiput(303.41,722.14)(0.509,-0.607){21}{\rule{0.123pt}{1.171pt}}
\multiput(300.34,724.57)(14.000,-14.569){2}{\rule{0.800pt}{0.586pt}}
\multiput(317.41,705.34)(0.509,-0.574){19}{\rule{0.123pt}{1.123pt}}
\multiput(314.34,707.67)(13.000,-12.669){2}{\rule{0.800pt}{0.562pt}}
\multiput(329.00,693.08)(0.581,-0.511){17}{\rule{1.133pt}{0.123pt}}
\multiput(329.00,693.34)(11.648,-12.000){2}{\rule{0.567pt}{0.800pt}}
\multiput(343.00,681.08)(0.581,-0.511){17}{\rule{1.133pt}{0.123pt}}
\multiput(343.00,681.34)(11.648,-12.000){2}{\rule{0.567pt}{0.800pt}}
\multiput(357.00,669.08)(0.589,-0.512){15}{\rule{1.145pt}{0.123pt}}
\multiput(357.00,669.34)(10.623,-11.000){2}{\rule{0.573pt}{0.800pt}}
\multiput(370.00,658.08)(0.710,-0.514){13}{\rule{1.320pt}{0.124pt}}
\multiput(370.00,658.34)(11.260,-10.000){2}{\rule{0.660pt}{0.800pt}}
\multiput(384.00,648.08)(0.710,-0.514){13}{\rule{1.320pt}{0.124pt}}
\multiput(384.00,648.34)(11.260,-10.000){2}{\rule{0.660pt}{0.800pt}}
\multiput(398.00,638.08)(0.737,-0.516){11}{\rule{1.356pt}{0.124pt}}
\multiput(398.00,638.34)(10.186,-9.000){2}{\rule{0.678pt}{0.800pt}}
\multiput(411.00,629.08)(0.800,-0.516){11}{\rule{1.444pt}{0.124pt}}
\multiput(411.00,629.34)(11.002,-9.000){2}{\rule{0.722pt}{0.800pt}}
\multiput(425.00,620.08)(0.800,-0.516){11}{\rule{1.444pt}{0.124pt}}
\multiput(425.00,620.34)(11.002,-9.000){2}{\rule{0.722pt}{0.800pt}}
\multiput(439.00,611.08)(0.737,-0.516){11}{\rule{1.356pt}{0.124pt}}
\multiput(439.00,611.34)(10.186,-9.000){2}{\rule{0.678pt}{0.800pt}}
\multiput(452.00,602.08)(0.920,-0.520){9}{\rule{1.600pt}{0.125pt}}
\multiput(452.00,602.34)(10.679,-8.000){2}{\rule{0.800pt}{0.800pt}}
\multiput(466.00,594.08)(1.088,-0.526){7}{\rule{1.800pt}{0.127pt}}
\multiput(466.00,594.34)(10.264,-7.000){2}{\rule{0.900pt}{0.800pt}}
\multiput(480.00,587.08)(0.737,-0.516){11}{\rule{1.356pt}{0.124pt}}
\multiput(480.00,587.34)(10.186,-9.000){2}{\rule{0.678pt}{0.800pt}}
\multiput(493.00,578.08)(1.088,-0.526){7}{\rule{1.800pt}{0.127pt}}
\multiput(493.00,578.34)(10.264,-7.000){2}{\rule{0.900pt}{0.800pt}}
\multiput(507.00,571.08)(0.920,-0.520){9}{\rule{1.600pt}{0.125pt}}
\multiput(507.00,571.34)(10.679,-8.000){2}{\rule{0.800pt}{0.800pt}}
\multiput(521.00,563.08)(1.000,-0.526){7}{\rule{1.686pt}{0.127pt}}
\multiput(521.00,563.34)(9.501,-7.000){2}{\rule{0.843pt}{0.800pt}}
\multiput(534.00,556.08)(1.088,-0.526){7}{\rule{1.800pt}{0.127pt}}
\multiput(534.00,556.34)(10.264,-7.000){2}{\rule{0.900pt}{0.800pt}}
\multiput(548.00,549.07)(1.355,-0.536){5}{\rule{2.067pt}{0.129pt}}
\multiput(548.00,549.34)(9.711,-6.000){2}{\rule{1.033pt}{0.800pt}}
\multiput(562.00,543.08)(1.000,-0.526){7}{\rule{1.686pt}{0.127pt}}
\multiput(562.00,543.34)(9.501,-7.000){2}{\rule{0.843pt}{0.800pt}}
\multiput(575.00,536.06)(1.936,-0.560){3}{\rule{2.440pt}{0.135pt}}
\multiput(575.00,536.34)(8.936,-5.000){2}{\rule{1.220pt}{0.800pt}}
\multiput(589.00,531.08)(1.000,-0.526){7}{\rule{1.686pt}{0.127pt}}
\multiput(589.00,531.34)(9.501,-7.000){2}{\rule{0.843pt}{0.800pt}}
\multiput(602.00,524.07)(1.355,-0.536){5}{\rule{2.067pt}{0.129pt}}
\multiput(602.00,524.34)(9.711,-6.000){2}{\rule{1.033pt}{0.800pt}}
\multiput(616.00,518.08)(1.088,-0.526){7}{\rule{1.800pt}{0.127pt}}
\multiput(616.00,518.34)(10.264,-7.000){2}{\rule{0.900pt}{0.800pt}}
\multiput(630.00,511.07)(1.244,-0.536){5}{\rule{1.933pt}{0.129pt}}
\multiput(630.00,511.34)(8.987,-6.000){2}{\rule{0.967pt}{0.800pt}}
\multiput(643.00,505.07)(1.355,-0.536){5}{\rule{2.067pt}{0.129pt}}
\multiput(643.00,505.34)(9.711,-6.000){2}{\rule{1.033pt}{0.800pt}}
\multiput(657.00,499.07)(1.355,-0.536){5}{\rule{2.067pt}{0.129pt}}
\multiput(657.00,499.34)(9.711,-6.000){2}{\rule{1.033pt}{0.800pt}}
\multiput(671.00,493.06)(1.768,-0.560){3}{\rule{2.280pt}{0.135pt}}
\multiput(671.00,493.34)(8.268,-5.000){2}{\rule{1.140pt}{0.800pt}}
\multiput(684.00,488.07)(1.355,-0.536){5}{\rule{2.067pt}{0.129pt}}
\multiput(684.00,488.34)(9.711,-6.000){2}{\rule{1.033pt}{0.800pt}}
\multiput(698.00,482.07)(1.355,-0.536){5}{\rule{2.067pt}{0.129pt}}
\multiput(698.00,482.34)(9.711,-6.000){2}{\rule{1.033pt}{0.800pt}}
\multiput(712.00,476.07)(1.244,-0.536){5}{\rule{1.933pt}{0.129pt}}
\multiput(712.00,476.34)(8.987,-6.000){2}{\rule{0.967pt}{0.800pt}}
\multiput(725.00,470.06)(1.936,-0.560){3}{\rule{2.440pt}{0.135pt}}
\multiput(725.00,470.34)(8.936,-5.000){2}{\rule{1.220pt}{0.800pt}}
\multiput(739.00,465.08)(1.088,-0.526){7}{\rule{1.800pt}{0.127pt}}
\multiput(739.00,465.34)(10.264,-7.000){2}{\rule{0.900pt}{0.800pt}}
\multiput(753.00,458.06)(1.768,-0.560){3}{\rule{2.280pt}{0.135pt}}
\multiput(753.00,458.34)(8.268,-5.000){2}{\rule{1.140pt}{0.800pt}}
\multiput(766.00,453.07)(1.355,-0.536){5}{\rule{2.067pt}{0.129pt}}
\multiput(766.00,453.34)(9.711,-6.000){2}{\rule{1.033pt}{0.800pt}}
\multiput(780.00,447.06)(1.936,-0.560){3}{\rule{2.440pt}{0.135pt}}
\multiput(780.00,447.34)(8.936,-5.000){2}{\rule{1.220pt}{0.800pt}}
\multiput(794.00,442.07)(1.244,-0.536){5}{\rule{1.933pt}{0.129pt}}
\multiput(794.00,442.34)(8.987,-6.000){2}{\rule{0.967pt}{0.800pt}}
\multiput(807.00,436.06)(1.936,-0.560){3}{\rule{2.440pt}{0.135pt}}
\multiput(807.00,436.34)(8.936,-5.000){2}{\rule{1.220pt}{0.800pt}}
\multiput(821.00,431.07)(1.355,-0.536){5}{\rule{2.067pt}{0.129pt}}
\multiput(821.00,431.34)(9.711,-6.000){2}{\rule{1.033pt}{0.800pt}}
\multiput(835.00,425.07)(1.244,-0.536){5}{\rule{1.933pt}{0.129pt}}
\multiput(835.00,425.34)(8.987,-6.000){2}{\rule{0.967pt}{0.800pt}}
\multiput(848.00,419.07)(1.355,-0.536){5}{\rule{2.067pt}{0.129pt}}
\multiput(848.00,419.34)(9.711,-6.000){2}{\rule{1.033pt}{0.800pt}}
\multiput(862.00,413.07)(1.355,-0.536){5}{\rule{2.067pt}{0.129pt}}
\multiput(862.00,413.34)(9.711,-6.000){2}{\rule{1.033pt}{0.800pt}}
\multiput(876.00,407.06)(1.768,-0.560){3}{\rule{2.280pt}{0.135pt}}
\multiput(876.00,407.34)(8.268,-5.000){2}{\rule{1.140pt}{0.800pt}}
\multiput(889.00,402.08)(1.088,-0.526){7}{\rule{1.800pt}{0.127pt}}
\multiput(889.00,402.34)(10.264,-7.000){2}{\rule{0.900pt}{0.800pt}}
\multiput(903.00,395.06)(1.936,-0.560){3}{\rule{2.440pt}{0.135pt}}
\multiput(903.00,395.34)(8.936,-5.000){2}{\rule{1.220pt}{0.800pt}}
\multiput(917.00,390.07)(1.244,-0.536){5}{\rule{1.933pt}{0.129pt}}
\multiput(917.00,390.34)(8.987,-6.000){2}{\rule{0.967pt}{0.800pt}}
\multiput(930.00,384.06)(1.936,-0.560){3}{\rule{2.440pt}{0.135pt}}
\multiput(930.00,384.34)(8.936,-5.000){2}{\rule{1.220pt}{0.800pt}}
\multiput(944.00,379.07)(1.355,-0.536){5}{\rule{2.067pt}{0.129pt}}
\multiput(944.00,379.34)(9.711,-6.000){2}{\rule{1.033pt}{0.800pt}}
\multiput(958.00,373.07)(1.244,-0.536){5}{\rule{1.933pt}{0.129pt}}
\multiput(958.00,373.34)(8.987,-6.000){2}{\rule{0.967pt}{0.800pt}}
\multiput(971.00,367.06)(1.936,-0.560){3}{\rule{2.440pt}{0.135pt}}
\multiput(971.00,367.34)(8.936,-5.000){2}{\rule{1.220pt}{0.800pt}}
\multiput(985.00,362.07)(1.355,-0.536){5}{\rule{2.067pt}{0.129pt}}
\multiput(985.00,362.34)(9.711,-6.000){2}{\rule{1.033pt}{0.800pt}}
\multiput(999.00,356.06)(1.768,-0.560){3}{\rule{2.280pt}{0.135pt}}
\multiput(999.00,356.34)(8.268,-5.000){2}{\rule{1.140pt}{0.800pt}}
\multiput(1012.00,351.06)(1.936,-0.560){3}{\rule{2.440pt}{0.135pt}}
\multiput(1012.00,351.34)(8.936,-5.000){2}{\rule{1.220pt}{0.800pt}}
\multiput(1026.00,346.06)(1.936,-0.560){3}{\rule{2.440pt}{0.135pt}}
\multiput(1026.00,346.34)(8.936,-5.000){2}{\rule{1.220pt}{0.800pt}}
\multiput(1040.00,341.08)(1.000,-0.526){7}{\rule{1.686pt}{0.127pt}}
\multiput(1040.00,341.34)(9.501,-7.000){2}{\rule{0.843pt}{0.800pt}}
\multiput(1053.00,334.06)(1.936,-0.560){3}{\rule{2.440pt}{0.135pt}}
\multiput(1053.00,334.34)(8.936,-5.000){2}{\rule{1.220pt}{0.800pt}}
\multiput(1067.00,329.07)(1.355,-0.536){5}{\rule{2.067pt}{0.129pt}}
\multiput(1067.00,329.34)(9.711,-6.000){2}{\rule{1.033pt}{0.800pt}}
\multiput(1081.00,323.06)(1.768,-0.560){3}{\rule{2.280pt}{0.135pt}}
\multiput(1081.00,323.34)(8.268,-5.000){2}{\rule{1.140pt}{0.800pt}}
\multiput(1094.00,318.06)(1.936,-0.560){3}{\rule{2.440pt}{0.135pt}}
\multiput(1094.00,318.34)(8.936,-5.000){2}{\rule{1.220pt}{0.800pt}}
\multiput(1108.00,313.06)(1.936,-0.560){3}{\rule{2.440pt}{0.135pt}}
\multiput(1108.00,313.34)(8.936,-5.000){2}{\rule{1.220pt}{0.800pt}}
\multiput(1122.00,308.06)(1.768,-0.560){3}{\rule{2.280pt}{0.135pt}}
\multiput(1122.00,308.34)(8.268,-5.000){2}{\rule{1.140pt}{0.800pt}}
\multiput(1135.00,303.06)(1.936,-0.560){3}{\rule{2.440pt}{0.135pt}}
\multiput(1135.00,303.34)(8.936,-5.000){2}{\rule{1.220pt}{0.800pt}}
\multiput(1149.00,298.07)(1.355,-0.536){5}{\rule{2.067pt}{0.129pt}}
\multiput(1149.00,298.34)(9.711,-6.000){2}{\rule{1.033pt}{0.800pt}}
\multiput(1163.00,292.07)(1.244,-0.536){5}{\rule{1.933pt}{0.129pt}}
\multiput(1163.00,292.34)(8.987,-6.000){2}{\rule{0.967pt}{0.800pt}}
\multiput(1176.00,286.06)(1.936,-0.560){3}{\rule{2.440pt}{0.135pt}}
\multiput(1176.00,286.34)(8.936,-5.000){2}{\rule{1.220pt}{0.800pt}}
\multiput(1190.00,281.06)(1.936,-0.560){3}{\rule{2.440pt}{0.135pt}}
\multiput(1190.00,281.34)(8.936,-5.000){2}{\rule{1.220pt}{0.800pt}}
\multiput(1204.00,276.06)(1.768,-0.560){3}{\rule{2.280pt}{0.135pt}}
\multiput(1204.00,276.34)(8.268,-5.000){2}{\rule{1.140pt}{0.800pt}}
\put(1217,269.34){\rule{3.000pt}{0.800pt}}
\multiput(1217.00,271.34)(7.773,-4.000){2}{\rule{1.500pt}{0.800pt}}
\multiput(1231.00,267.07)(1.355,-0.536){5}{\rule{2.067pt}{0.129pt}}
\multiput(1231.00,267.34)(9.711,-6.000){2}{\rule{1.033pt}{0.800pt}}
\put(1245,259.34){\rule{2.800pt}{0.800pt}}
\multiput(1245.00,261.34)(7.188,-4.000){2}{\rule{1.400pt}{0.800pt}}
\put(1258,255.34){\rule{3.000pt}{0.800pt}}
\multiput(1258.00,257.34)(7.773,-4.000){2}{\rule{1.500pt}{0.800pt}}
\multiput(1272.00,253.06)(1.768,-0.560){3}{\rule{2.280pt}{0.135pt}}
\multiput(1272.00,253.34)(8.268,-5.000){2}{\rule{1.140pt}{0.800pt}}
\put(1285,246.34){\rule{3.000pt}{0.800pt}}
\multiput(1285.00,248.34)(7.773,-4.000){2}{\rule{1.500pt}{0.800pt}}
\multiput(1299.00,244.07)(1.355,-0.536){5}{\rule{2.067pt}{0.129pt}}
\multiput(1299.00,244.34)(9.711,-6.000){2}{\rule{1.033pt}{0.800pt}}
\multiput(1313.00,238.07)(1.244,-0.536){5}{\rule{1.933pt}{0.129pt}}
\multiput(1313.00,238.34)(8.987,-6.000){2}{\rule{0.967pt}{0.800pt}}
\multiput(1326.00,232.06)(1.936,-0.560){3}{\rule{2.440pt}{0.135pt}}
\multiput(1326.00,232.34)(8.936,-5.000){2}{\rule{1.220pt}{0.800pt}}
\put(1340,225.34){\rule{3.000pt}{0.800pt}}
\multiput(1340.00,227.34)(7.773,-4.000){2}{\rule{1.500pt}{0.800pt}}
\put(1354,221.34){\rule{2.800pt}{0.800pt}}
\multiput(1354.00,223.34)(7.188,-4.000){2}{\rule{1.400pt}{0.800pt}}
\sbox{\plotpoint}{\rule[-0.200pt]{0.400pt}{0.400pt}}%
\multiput(261,1146)(2.906,-20.551){5}{\usebox{\plotpoint}}
\multiput(275,1047)(3.412,-20.473){4}{\usebox{\plotpoint}}
\multiput(288,969)(6.293,-19.778){2}{\usebox{\plotpoint}}
\multiput(302,925)(9.554,-18.426){2}{\usebox{\plotpoint}}
\put(324.61,886.74){\usebox{\plotpoint}}
\put(336.50,869.74){\usebox{\plotpoint}}
\put(348.75,853.02){\usebox{\plotpoint}}
\put(361.72,836.82){\usebox{\plotpoint}}
\put(373.71,819.91){\usebox{\plotpoint}}
\put(384.63,802.28){\usebox{\plotpoint}}
\multiput(398,787)(11.720,-17.130){2}{\usebox{\plotpoint}}
\multiput(411,768)(16.320,-12.823){0}{\usebox{\plotpoint}}
\multiput(425,757)(13.194,-16.022){2}{\usebox{\plotpoint}}
\multiput(439,740)(15.844,-13.407){0}{\usebox{\plotpoint}}
\put(454.21,726.16){\usebox{\plotpoint}}
\multiput(466,711)(9.282,-18.564){2}{\usebox{\plotpoint}}
\put(487.70,674.11){\usebox{\plotpoint}}
\put(501.05,658.22){\usebox{\plotpoint}}
\put(517.19,645.90){\usebox{\plotpoint}}
\multiput(521,644)(7.227,-19.457){2}{\usebox{\plotpoint}}
\multiput(534,609)(5.812,-19.925){2}{\usebox{\plotpoint}}
\put(556.93,569.29){\usebox{\plotpoint}}
\multiput(562,574)(5.127,-20.112){2}{\usebox{\plotpoint}}
\put(576.85,524.99){\usebox{\plotpoint}}
\multiput(589,538)(6.563,-19.690){2}{\usebox{\plotpoint}}
\multiput(602,499)(5.120,20.114){3}{\usebox{\plotpoint}}
\multiput(616,554)(4.792,-20.195){3}{\usebox{\plotpoint}}
\multiput(630,495)(3.412,20.473){4}{\usebox{\plotpoint}}
\multiput(643,573)(20.756,0.000){0}{\usebox{\plotpoint}}
\put(657.41,572.83){\usebox{\plotpoint}}
\put(674.76,571.63){\usebox{\plotpoint}}
\put(690.04,583.86){\usebox{\plotpoint}}
\multiput(698,585)(7.708,-19.271){2}{\usebox{\plotpoint}}
\put(721.34,563.65){\usebox{\plotpoint}}
\put(738.44,564.20){\usebox{\plotpoint}}
\put(748.01,545.97){\usebox{\plotpoint}}
\put(762.61,536.00){\usebox{\plotpoint}}
\put(774.48,551.15){\usebox{\plotpoint}}
\put(786.23,553.88){\usebox{\plotpoint}}
\put(802.56,547.64){\usebox{\plotpoint}}
\put(817.24,536.56){\usebox{\plotpoint}}
\multiput(821,532)(6.857,-19.590){2}{\usebox{\plotpoint}}
\multiput(835,492)(4.693,20.218){3}{\usebox{\plotpoint}}
\multiput(848,548)(19.077,-8.176){0}{\usebox{\plotpoint}}
\multiput(862,542)(11.513,-17.270){2}{\usebox{\plotpoint}}
\multiput(876,521)(4.466,-20.269){2}{\usebox{\plotpoint}}
\multiput(889,462)(8.319,19.015){2}{\usebox{\plotpoint}}
\put(908.10,499.47){\usebox{\plotpoint}}
\put(924.37,506.73){\usebox{\plotpoint}}
\put(941.69,495.81){\usebox{\plotpoint}}
\multiput(944,494)(3.156,-20.514){4}{\usebox{\plotpoint}}
\put(963.74,409.63){\usebox{\plotpoint}}
\multiput(971,418)(5.396,20.042){3}{\usebox{\plotpoint}}
\put(995.37,455.18){\usebox{\plotpoint}}
\put(1007.77,461.46){\usebox{\plotpoint}}
\put(1019.40,478.63){\usebox{\plotpoint}}
\put(1033.12,493.58){\usebox{\plotpoint}}
\multiput(1040,498)(2.729,-20.575){5}{\usebox{\plotpoint}}
\put(1063.38,414.09){\usebox{\plotpoint}}
\multiput(1067,419)(6.426,-19.736){2}{\usebox{\plotpoint}}
\multiput(1081,376)(8.027,-19.141){2}{\usebox{\plotpoint}}
\multiput(1094,345)(15.759,13.508){0}{\usebox{\plotpoint}}
\multiput(1108,357)(3.962,20.374){4}{\usebox{\plotpoint}}
\multiput(1122,429)(7.413,-19.387){2}{\usebox{\plotpoint}}
\put(1148.95,389.02){\usebox{\plotpoint}}
\multiput(1149,389)(1.684,-20.687){8}{\usebox{\plotpoint}}
\multiput(1163,217)(4.615,20.236){3}{\usebox{\plotpoint}}
\multiput(1176,274)(4.503,-20.261){3}{\usebox{\plotpoint}}
\multiput(1190,211)(5.034,20.136){3}{\usebox{\plotpoint}}
\put(1216.30,283.09){\usebox{\plotpoint}}
\put(1224.86,301.97){\usebox{\plotpoint}}
\put(1236.42,315.61){\usebox{\plotpoint}}
\put(1250.12,326.03){\usebox{\plotpoint}}
\multiput(1258,343)(8.543,18.916){2}{\usebox{\plotpoint}}
\put(1280.66,370.00){\usebox{\plotpoint}}
\put(1295.16,355.67){\usebox{\plotpoint}}
\multiput(1299,351)(2.425,-20.613){6}{\usebox{\plotpoint}}
\put(1319.76,250.20){\usebox{\plotpoint}}
\put(1328.31,265.35){\usebox{\plotpoint}}
\put(1344.36,252.33){\usebox{\plotpoint}}
\multiput(1354,242)(2.436,-20.612){6}{\usebox{\plotpoint}}
\put(1367,132){\usebox{\plotpoint}}
\end{picture}
\end{center}
\caption[]{Values of $H_L$ and $\vartheta_H$ as functions of number of
iterations $N_{iter}$ of the covariant gauge-fixing algorithm obtained
extremizing $H_L$ for a configuration of $SU(3)$ with $a=0.3$, $V=4^4$ and
$\Lambda(x)=0$.\label{grf1}}
\end{figure}
In fig. \ref{grf1} we plot $H_L$ and $\vartheta_H$ as a function of the
number of gauge-fixing sweeps for a configuration of
$SU(3)$ with $a=0.3$ and $\Lambda(x)=0$. It is easy to see that both $H_L$ and
$\vartheta_H$ go to values consistent with the precision required for gauge-fixing ($\leq 10^{-5}$). For all configurations we generated the behaviour of $H_L$ and
$\vartheta_H$ is consistent with that shown in fig.\ref{grf1}. 
For $\Lambda(x)=0$ we fixed the gauge condition $\Delta(x)
\leq 10^{-5}$ for each configuration also using
the old Landau gauge-fixing algorithm. For the two configurations obtained with
the two algorithms, we measured the discretization of $F(\Omega)$
\be\label{gribovvecchio}
F_L(\Omega)=-\frac{1}{V}\sum_{x,\mu}\mbox{\rm ReTr}
U^\Omega\m(x)
\ee
as a measure of the distance between the two configurations. The
values are the same with a precision of $10^{-5}$, consistent with the precision 
required to minimize $H_L$ .\\
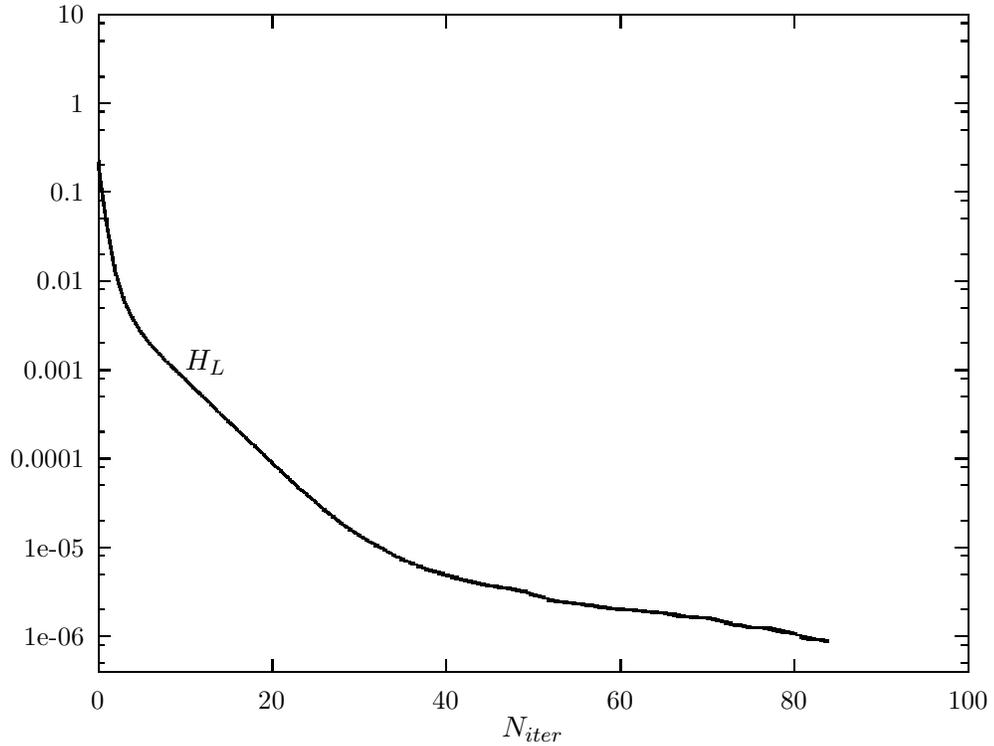
\begin{figure}[htb]   
\begin{center}
% GNUPLOT: LaTeX picture
\setlength{\unitlength}{0.240900pt}
\ifx\plotpoint\undefined\newsavebox{\plotpoint}\fi
\sbox{\plotpoint}{\rule[-0.200pt]{0.400pt}{0.400pt}}%
\begin{picture}(1650,1169)(0,0)
\font\gnuplot=cmr10 at 10pt
\gnuplot
\sbox{\plotpoint}{\rule[-0.200pt]{0.400pt}{0.400pt}}%
\put(220.0,113.0){\rule[-0.200pt]{0.400pt}{248.850pt}}
\put(220.0,127.0){\rule[-0.200pt]{2.409pt}{0.400pt}}
\put(1576.0,127.0){\rule[-0.200pt]{2.409pt}{0.400pt}}
\put(220.0,155.0){\rule[-0.200pt]{2.409pt}{0.400pt}}
\put(1576.0,155.0){\rule[-0.200pt]{2.409pt}{0.400pt}}
\put(220.0,169.0){\rule[-0.200pt]{4.818pt}{0.400pt}}
\put(198,169){\makebox(0,0)[r]{1e-06}}
\put(1566.0,169.0){\rule[-0.200pt]{4.818pt}{0.400pt}}
\put(220.0,211.0){\rule[-0.200pt]{2.409pt}{0.400pt}}
\put(1576.0,211.0){\rule[-0.200pt]{2.409pt}{0.400pt}}
\put(220.0,266.0){\rule[-0.200pt]{2.409pt}{0.400pt}}
\put(1576.0,266.0){\rule[-0.200pt]{2.409pt}{0.400pt}}
\put(220.0,295.0){\rule[-0.200pt]{2.409pt}{0.400pt}}
\put(1576.0,295.0){\rule[-0.200pt]{2.409pt}{0.400pt}}
\put(220.0,308.0){\rule[-0.200pt]{4.818pt}{0.400pt}}
\put(198,308){\makebox(0,0)[r]{1e-05}}
\put(1566.0,308.0){\rule[-0.200pt]{4.818pt}{0.400pt}}
\put(220.0,350.0){\rule[-0.200pt]{2.409pt}{0.400pt}}
\put(1576.0,350.0){\rule[-0.200pt]{2.409pt}{0.400pt}}
\put(220.0,406.0){\rule[-0.200pt]{2.409pt}{0.400pt}}
\put(1576.0,406.0){\rule[-0.200pt]{2.409pt}{0.400pt}}
\put(220.0,434.0){\rule[-0.200pt]{2.409pt}{0.400pt}}
\put(1576.0,434.0){\rule[-0.200pt]{2.409pt}{0.400pt}}
\put(220.0,448.0){\rule[-0.200pt]{4.818pt}{0.400pt}}
\put(198,448){\makebox(0,0)[r]{0.0001}}
\put(1566.0,448.0){\rule[-0.200pt]{4.818pt}{0.400pt}}
\put(220.0,490.0){\rule[-0.200pt]{2.409pt}{0.400pt}}
\put(1576.0,490.0){\rule[-0.200pt]{2.409pt}{0.400pt}}
\put(220.0,545.0){\rule[-0.200pt]{2.409pt}{0.400pt}}
\put(1576.0,545.0){\rule[-0.200pt]{2.409pt}{0.400pt}}
\put(220.0,574.0){\rule[-0.200pt]{2.409pt}{0.400pt}}
\put(1576.0,574.0){\rule[-0.200pt]{2.409pt}{0.400pt}}
\put(220.0,587.0){\rule[-0.200pt]{4.818pt}{0.400pt}}
\put(198,587){\makebox(0,0)[r]{0.001}}
\put(1566.0,587.0){\rule[-0.200pt]{4.818pt}{0.400pt}}
\put(220.0,630.0){\rule[-0.200pt]{2.409pt}{0.400pt}}
\put(1576.0,630.0){\rule[-0.200pt]{2.409pt}{0.400pt}}
\put(220.0,685.0){\rule[-0.200pt]{2.409pt}{0.400pt}}
\put(1576.0,685.0){\rule[-0.200pt]{2.409pt}{0.400pt}}
\put(220.0,714.0){\rule[-0.200pt]{2.409pt}{0.400pt}}
\put(1576.0,714.0){\rule[-0.200pt]{2.409pt}{0.400pt}}
\put(220.0,727.0){\rule[-0.200pt]{4.818pt}{0.400pt}}
\put(198,727){\makebox(0,0)[r]{0.01}}
\put(1566.0,727.0){\rule[-0.200pt]{4.818pt}{0.400pt}}
\put(220.0,769.0){\rule[-0.200pt]{2.409pt}{0.400pt}}
\put(1576.0,769.0){\rule[-0.200pt]{2.409pt}{0.400pt}}
\put(220.0,825.0){\rule[-0.200pt]{2.409pt}{0.400pt}}
\put(1576.0,825.0){\rule[-0.200pt]{2.409pt}{0.400pt}}
\put(220.0,853.0){\rule[-0.200pt]{2.409pt}{0.400pt}}
\put(1576.0,853.0){\rule[-0.200pt]{2.409pt}{0.400pt}}
\put(220.0,867.0){\rule[-0.200pt]{4.818pt}{0.400pt}}
\put(198,867){\makebox(0,0)[r]{0.1}}
\put(1566.0,867.0){\rule[-0.200pt]{4.818pt}{0.400pt}}
\put(220.0,909.0){\rule[-0.200pt]{2.409pt}{0.400pt}}
\put(1576.0,909.0){\rule[-0.200pt]{2.409pt}{0.400pt}}
\put(220.0,964.0){\rule[-0.200pt]{2.409pt}{0.400pt}}
\put(1576.0,964.0){\rule[-0.200pt]{2.409pt}{0.400pt}}
\put(220.0,993.0){\rule[-0.200pt]{2.409pt}{0.400pt}}
\put(1576.0,993.0){\rule[-0.200pt]{2.409pt}{0.400pt}}
\put(220.0,1006.0){\rule[-0.200pt]{4.818pt}{0.400pt}}
\put(198,1006){\makebox(0,0)[r]{1}}
\put(1566.0,1006.0){\rule[-0.200pt]{4.818pt}{0.400pt}}
\put(220.0,1048.0){\rule[-0.200pt]{2.409pt}{0.400pt}}
\put(1576.0,1048.0){\rule[-0.200pt]{2.409pt}{0.400pt}}
\put(220.0,1104.0){\rule[-0.200pt]{2.409pt}{0.400pt}}
\put(1576.0,1104.0){\rule[-0.200pt]{2.409pt}{0.400pt}}
\put(220.0,1132.0){\rule[-0.200pt]{2.409pt}{0.400pt}}
\put(1576.0,1132.0){\rule[-0.200pt]{2.409pt}{0.400pt}}
\put(220.0,1146.0){\rule[-0.200pt]{4.818pt}{0.400pt}}
\put(198,1146){\makebox(0,0)[r]{10}}
\put(1566.0,1146.0){\rule[-0.200pt]{4.818pt}{0.400pt}}
\put(220.0,113.0){\rule[-0.200pt]{0.400pt}{4.818pt}}
\put(220,68){\makebox(0,0){0}}
\put(220.0,1126.0){\rule[-0.200pt]{0.400pt}{4.818pt}}
\put(493.0,113.0){\rule[-0.200pt]{0.400pt}{4.818pt}}
\put(493,68){\makebox(0,0){20}}
\put(493.0,1126.0){\rule[-0.200pt]{0.400pt}{4.818pt}}
\put(766.0,113.0){\rule[-0.200pt]{0.400pt}{4.818pt}}
\put(766,68){\makebox(0,0){40}}
\put(766.0,1126.0){\rule[-0.200pt]{0.400pt}{4.818pt}}
\put(1040.0,113.0){\rule[-0.200pt]{0.400pt}{4.818pt}}
\put(1040,68){\makebox(0,0){60}}
\put(1040.0,1126.0){\rule[-0.200pt]{0.400pt}{4.818pt}}
\put(1313.0,113.0){\rule[-0.200pt]{0.400pt}{4.818pt}}
\put(1313,68){\makebox(0,0){80}}
\put(1313.0,1126.0){\rule[-0.200pt]{0.400pt}{4.818pt}}
\put(1586.0,113.0){\rule[-0.200pt]{0.400pt}{4.818pt}}
\put(1586,68){\makebox(0,0){100}}
\put(1586.0,1126.0){\rule[-0.200pt]{0.400pt}{4.818pt}}
\put(220.0,113.0){\rule[-0.200pt]{329.069pt}{0.400pt}}
\put(1586.0,113.0){\rule[-0.200pt]{0.400pt}{248.850pt}}
\put(220.0,1146.0){\rule[-0.200pt]{329.069pt}{0.400pt}}
\put(45,629){\makebox(0,0){$ $}}
\put(903,23){\makebox(0,0){$N_{iter}$}}
\put(357,599){\makebox(0,0)[l]{$H_L$}}
\put(220.0,113.0){\rule[-0.200pt]{0.400pt}{248.850pt}}
\sbox{\plotpoint}{\rule[-0.400pt]{0.800pt}{0.800pt}}%
\put(220,914){\usebox{\plotpoint}}
\multiput(221.41,889.45)(0.509,-3.772){21}{\rule{0.123pt}{5.914pt}}
\multiput(218.34,901.72)(14.000,-87.725){2}{\rule{0.800pt}{2.957pt}}
\multiput(235.41,794.78)(0.509,-2.932){19}{\rule{0.123pt}{4.631pt}}
\multiput(232.34,804.39)(13.000,-62.389){2}{\rule{0.800pt}{2.315pt}}
\multiput(248.41,730.26)(0.509,-1.713){21}{\rule{0.123pt}{2.829pt}}
\multiput(245.34,736.13)(14.000,-40.129){2}{\rule{0.800pt}{1.414pt}}
\multiput(262.41,688.29)(0.509,-1.065){21}{\rule{0.123pt}{1.857pt}}
\multiput(259.34,692.15)(14.000,-25.145){2}{\rule{0.800pt}{0.929pt}}
\multiput(276.41,660.81)(0.509,-0.823){19}{\rule{0.123pt}{1.492pt}}
\multiput(273.34,663.90)(13.000,-17.903){2}{\rule{0.800pt}{0.746pt}}
\multiput(289.41,640.90)(0.509,-0.645){21}{\rule{0.123pt}{1.229pt}}
\multiput(286.34,643.45)(14.000,-15.450){2}{\rule{0.800pt}{0.614pt}}
\multiput(303.41,623.61)(0.509,-0.531){21}{\rule{0.123pt}{1.057pt}}
\multiput(300.34,625.81)(14.000,-12.806){2}{\rule{0.800pt}{0.529pt}}
\multiput(317.41,608.59)(0.509,-0.533){19}{\rule{0.123pt}{1.062pt}}
\multiput(314.34,610.80)(13.000,-11.797){2}{\rule{0.800pt}{0.531pt}}
\multiput(329.00,597.09)(0.492,-0.509){21}{\rule{1.000pt}{0.123pt}}
\multiput(329.00,597.34)(11.924,-14.000){2}{\rule{0.500pt}{0.800pt}}
\multiput(343.00,583.08)(0.533,-0.509){19}{\rule{1.062pt}{0.123pt}}
\multiput(343.00,583.34)(11.797,-13.000){2}{\rule{0.531pt}{0.800pt}}
\multiput(357.00,570.08)(0.492,-0.509){19}{\rule{1.000pt}{0.123pt}}
\multiput(357.00,570.34)(10.924,-13.000){2}{\rule{0.500pt}{0.800pt}}
\multiput(370.00,557.08)(0.533,-0.509){19}{\rule{1.062pt}{0.123pt}}
\multiput(370.00,557.34)(11.797,-13.000){2}{\rule{0.531pt}{0.800pt}}
\multiput(384.00,544.08)(0.533,-0.509){19}{\rule{1.062pt}{0.123pt}}
\multiput(384.00,544.34)(11.797,-13.000){2}{\rule{0.531pt}{0.800pt}}
\multiput(399.41,528.59)(0.509,-0.533){19}{\rule{0.123pt}{1.062pt}}
\multiput(396.34,530.80)(13.000,-11.797){2}{\rule{0.800pt}{0.531pt}}
\multiput(411.00,517.08)(0.533,-0.509){19}{\rule{1.062pt}{0.123pt}}
\multiput(411.00,517.34)(11.797,-13.000){2}{\rule{0.531pt}{0.800pt}}
\multiput(425.00,504.08)(0.533,-0.509){19}{\rule{1.062pt}{0.123pt}}
\multiput(425.00,504.34)(11.797,-13.000){2}{\rule{0.531pt}{0.800pt}}
\multiput(439.00,491.08)(0.492,-0.509){19}{\rule{1.000pt}{0.123pt}}
\multiput(439.00,491.34)(10.924,-13.000){2}{\rule{0.500pt}{0.800pt}}
\multiput(452.00,478.08)(0.533,-0.509){19}{\rule{1.062pt}{0.123pt}}
\multiput(452.00,478.34)(11.797,-13.000){2}{\rule{0.531pt}{0.800pt}}
\multiput(466.00,465.08)(0.533,-0.509){19}{\rule{1.062pt}{0.123pt}}
\multiput(466.00,465.34)(11.797,-13.000){2}{\rule{0.531pt}{0.800pt}}
\multiput(480.00,452.08)(0.492,-0.509){19}{\rule{1.000pt}{0.123pt}}
\multiput(480.00,452.34)(10.924,-13.000){2}{\rule{0.500pt}{0.800pt}}
\multiput(493.00,439.08)(0.533,-0.509){19}{\rule{1.062pt}{0.123pt}}
\multiput(493.00,439.34)(11.797,-13.000){2}{\rule{0.531pt}{0.800pt}}
\multiput(507.00,426.08)(0.533,-0.509){19}{\rule{1.062pt}{0.123pt}}
\multiput(507.00,426.34)(11.797,-13.000){2}{\rule{0.531pt}{0.800pt}}
\multiput(521.00,413.08)(0.492,-0.509){19}{\rule{1.000pt}{0.123pt}}
\multiput(521.00,413.34)(10.924,-13.000){2}{\rule{0.500pt}{0.800pt}}
\multiput(534.00,400.08)(0.639,-0.512){15}{\rule{1.218pt}{0.123pt}}
\multiput(534.00,400.34)(11.472,-11.000){2}{\rule{0.609pt}{0.800pt}}
\multiput(548.00,389.08)(0.581,-0.511){17}{\rule{1.133pt}{0.123pt}}
\multiput(548.00,389.34)(11.648,-12.000){2}{\rule{0.567pt}{0.800pt}}
\multiput(562.00,377.08)(0.536,-0.511){17}{\rule{1.067pt}{0.123pt}}
\multiput(562.00,377.34)(10.786,-12.000){2}{\rule{0.533pt}{0.800pt}}
\multiput(575.00,365.08)(0.639,-0.512){15}{\rule{1.218pt}{0.123pt}}
\multiput(575.00,365.34)(11.472,-11.000){2}{\rule{0.609pt}{0.800pt}}
\multiput(589.00,354.08)(0.654,-0.514){13}{\rule{1.240pt}{0.124pt}}
\multiput(589.00,354.34)(10.426,-10.000){2}{\rule{0.620pt}{0.800pt}}
\multiput(602.00,344.08)(0.710,-0.514){13}{\rule{1.320pt}{0.124pt}}
\multiput(602.00,344.34)(11.260,-10.000){2}{\rule{0.660pt}{0.800pt}}
\multiput(616.00,334.08)(0.800,-0.516){11}{\rule{1.444pt}{0.124pt}}
\multiput(616.00,334.34)(11.002,-9.000){2}{\rule{0.722pt}{0.800pt}}
\multiput(630.00,325.08)(0.847,-0.520){9}{\rule{1.500pt}{0.125pt}}
\multiput(630.00,325.34)(9.887,-8.000){2}{\rule{0.750pt}{0.800pt}}
\multiput(643.00,317.08)(0.920,-0.520){9}{\rule{1.600pt}{0.125pt}}
\multiput(643.00,317.34)(10.679,-8.000){2}{\rule{0.800pt}{0.800pt}}
\multiput(657.00,309.08)(1.088,-0.526){7}{\rule{1.800pt}{0.127pt}}
\multiput(657.00,309.34)(10.264,-7.000){2}{\rule{0.900pt}{0.800pt}}
\multiput(671.00,302.08)(0.847,-0.520){9}{\rule{1.500pt}{0.125pt}}
\multiput(671.00,302.34)(9.887,-8.000){2}{\rule{0.750pt}{0.800pt}}
\multiput(684.00,294.08)(1.088,-0.526){7}{\rule{1.800pt}{0.127pt}}
\multiput(684.00,294.34)(10.264,-7.000){2}{\rule{0.900pt}{0.800pt}}
\multiput(698.00,287.07)(1.355,-0.536){5}{\rule{2.067pt}{0.129pt}}
\multiput(698.00,287.34)(9.711,-6.000){2}{\rule{1.033pt}{0.800pt}}
\multiput(712.00,281.06)(1.768,-0.560){3}{\rule{2.280pt}{0.135pt}}
\multiput(712.00,281.34)(8.268,-5.000){2}{\rule{1.140pt}{0.800pt}}
\multiput(725.00,276.06)(1.936,-0.560){3}{\rule{2.440pt}{0.135pt}}
\multiput(725.00,276.34)(8.936,-5.000){2}{\rule{1.220pt}{0.800pt}}
\put(739,269.34){\rule{3.000pt}{0.800pt}}
\multiput(739.00,271.34)(7.773,-4.000){2}{\rule{1.500pt}{0.800pt}}
\put(753,265.34){\rule{2.800pt}{0.800pt}}
\multiput(753.00,267.34)(7.188,-4.000){2}{\rule{1.400pt}{0.800pt}}
\put(766,261.34){\rule{3.000pt}{0.800pt}}
\multiput(766.00,263.34)(7.773,-4.000){2}{\rule{1.500pt}{0.800pt}}
\put(780,257.34){\rule{3.000pt}{0.800pt}}
\multiput(780.00,259.34)(7.773,-4.000){2}{\rule{1.500pt}{0.800pt}}
\put(794,253.84){\rule{3.132pt}{0.800pt}}
\multiput(794.00,255.34)(6.500,-3.000){2}{\rule{1.566pt}{0.800pt}}
\put(807,250.84){\rule{3.373pt}{0.800pt}}
\multiput(807.00,252.34)(7.000,-3.000){2}{\rule{1.686pt}{0.800pt}}
\put(821,247.84){\rule{3.373pt}{0.800pt}}
\multiput(821.00,249.34)(7.000,-3.000){2}{\rule{1.686pt}{0.800pt}}
\put(835,245.34){\rule{3.132pt}{0.800pt}}
\multiput(835.00,246.34)(6.500,-2.000){2}{\rule{1.566pt}{0.800pt}}
\put(848,243.34){\rule{3.373pt}{0.800pt}}
\multiput(848.00,244.34)(7.000,-2.000){2}{\rule{1.686pt}{0.800pt}}
\put(862,240.84){\rule{3.373pt}{0.800pt}}
\multiput(862.00,242.34)(7.000,-3.000){2}{\rule{1.686pt}{0.800pt}}
\put(876,238.34){\rule{3.132pt}{0.800pt}}
\multiput(876.00,239.34)(6.500,-2.000){2}{\rule{1.566pt}{0.800pt}}
\multiput(889.00,237.06)(1.936,-0.560){3}{\rule{2.440pt}{0.135pt}}
\multiput(889.00,237.34)(8.936,-5.000){2}{\rule{1.220pt}{0.800pt}}
\put(903,230.34){\rule{3.000pt}{0.800pt}}
\multiput(903.00,232.34)(7.773,-4.000){2}{\rule{1.500pt}{0.800pt}}
\multiput(917.00,228.06)(1.768,-0.560){3}{\rule{2.280pt}{0.135pt}}
\multiput(917.00,228.34)(8.268,-5.000){2}{\rule{1.140pt}{0.800pt}}
\put(930,222.34){\rule{3.373pt}{0.800pt}}
\multiput(930.00,223.34)(7.000,-2.000){2}{\rule{1.686pt}{0.800pt}}
\put(944,220.34){\rule{3.373pt}{0.800pt}}
\multiput(944.00,221.34)(7.000,-2.000){2}{\rule{1.686pt}{0.800pt}}
\put(958,218.84){\rule{3.132pt}{0.800pt}}
\multiput(958.00,219.34)(6.500,-1.000){2}{\rule{1.566pt}{0.800pt}}
\put(971,217.34){\rule{3.373pt}{0.800pt}}
\multiput(971.00,218.34)(7.000,-2.000){2}{\rule{1.686pt}{0.800pt}}
\put(985,215.34){\rule{3.373pt}{0.800pt}}
\multiput(985.00,216.34)(7.000,-2.000){2}{\rule{1.686pt}{0.800pt}}
\put(999,213.34){\rule{3.132pt}{0.800pt}}
\multiput(999.00,214.34)(6.500,-2.000){2}{\rule{1.566pt}{0.800pt}}
\put(1012,211.34){\rule{3.373pt}{0.800pt}}
\multiput(1012.00,212.34)(7.000,-2.000){2}{\rule{1.686pt}{0.800pt}}
\put(1026,209.84){\rule{3.373pt}{0.800pt}}
\multiput(1026.00,210.34)(7.000,-1.000){2}{\rule{1.686pt}{0.800pt}}
\put(1040,208.84){\rule{3.132pt}{0.800pt}}
\multiput(1040.00,209.34)(6.500,-1.000){2}{\rule{1.566pt}{0.800pt}}
\put(1053,207.84){\rule{3.373pt}{0.800pt}}
\multiput(1053.00,208.34)(7.000,-1.000){2}{\rule{1.686pt}{0.800pt}}
\put(1067,206.34){\rule{3.373pt}{0.800pt}}
\multiput(1067.00,207.34)(7.000,-2.000){2}{\rule{1.686pt}{0.800pt}}
\put(1081,204.84){\rule{3.132pt}{0.800pt}}
\multiput(1081.00,205.34)(6.500,-1.000){2}{\rule{1.566pt}{0.800pt}}
\put(1094,203.84){\rule{3.373pt}{0.800pt}}
\multiput(1094.00,204.34)(7.000,-1.000){2}{\rule{1.686pt}{0.800pt}}
\put(1108,202.34){\rule{3.373pt}{0.800pt}}
\multiput(1108.00,203.34)(7.000,-2.000){2}{\rule{1.686pt}{0.800pt}}
\put(1122,199.84){\rule{3.132pt}{0.800pt}}
\multiput(1122.00,201.34)(6.500,-3.000){2}{\rule{1.566pt}{0.800pt}}
\put(1135,197.84){\rule{3.373pt}{0.800pt}}
\multiput(1135.00,198.34)(7.000,-1.000){2}{\rule{1.686pt}{0.800pt}}
\put(1149,196.84){\rule{3.373pt}{0.800pt}}
\multiput(1149.00,197.34)(7.000,-1.000){2}{\rule{1.686pt}{0.800pt}}
\put(1163,195.84){\rule{3.132pt}{0.800pt}}
\multiput(1163.00,196.34)(6.500,-1.000){2}{\rule{1.566pt}{0.800pt}}
\put(1176,194.34){\rule{3.373pt}{0.800pt}}
\multiput(1176.00,195.34)(7.000,-2.000){2}{\rule{1.686pt}{0.800pt}}
\put(1190,191.34){\rule{3.000pt}{0.800pt}}
\multiput(1190.00,193.34)(7.773,-4.000){2}{\rule{1.500pt}{0.800pt}}
\put(1204,187.34){\rule{2.800pt}{0.800pt}}
\multiput(1204.00,189.34)(7.188,-4.000){2}{\rule{1.400pt}{0.800pt}}
\put(1217,184.34){\rule{3.373pt}{0.800pt}}
\multiput(1217.00,185.34)(7.000,-2.000){2}{\rule{1.686pt}{0.800pt}}
\put(1231,181.84){\rule{3.373pt}{0.800pt}}
\multiput(1231.00,183.34)(7.000,-3.000){2}{\rule{1.686pt}{0.800pt}}
\put(1258,179.84){\rule{3.373pt}{0.800pt}}
\multiput(1258.00,180.34)(7.000,-1.000){2}{\rule{1.686pt}{0.800pt}}
\put(1272,177.84){\rule{3.132pt}{0.800pt}}
\multiput(1272.00,179.34)(6.500,-3.000){2}{\rule{1.566pt}{0.800pt}}
\put(1285,174.84){\rule{3.373pt}{0.800pt}}
\multiput(1285.00,176.34)(7.000,-3.000){2}{\rule{1.686pt}{0.800pt}}
\put(1299,172.34){\rule{3.373pt}{0.800pt}}
\multiput(1299.00,173.34)(7.000,-2.000){2}{\rule{1.686pt}{0.800pt}}
\multiput(1313.00,171.07)(1.244,-0.536){5}{\rule{1.933pt}{0.129pt}}
\multiput(1313.00,171.34)(8.987,-6.000){2}{\rule{0.967pt}{0.800pt}}
\put(1326,163.84){\rule{3.373pt}{0.800pt}}
\multiput(1326.00,165.34)(7.000,-3.000){2}{\rule{1.686pt}{0.800pt}}
\put(1340,161.84){\rule{3.373pt}{0.800pt}}
\multiput(1340.00,162.34)(7.000,-1.000){2}{\rule{1.686pt}{0.800pt}}
\put(1354,159.84){\rule{3.132pt}{0.800pt}}
\multiput(1354.00,161.34)(6.500,-3.000){2}{\rule{1.566pt}{0.800pt}}
\put(1245.0,182.0){\rule[-0.400pt]{3.132pt}{0.800pt}}
\end{picture}
\end{center}
\caption[]{Values of $H_L$ and $\vartheta_H$ as functions of number of
iterations $N_{iter}$ of the covariant gauge-fixing algorithm obtained
extremizing $H_L$ for a configuration of $SU(2)$ with $a=0.1$, $V=4^4$ and
$\Lambda(x)\neq 0$.\label{grf2}}
\end{figure}
In fig. \ref{grf2} we plot $H_L$ as a function of the number of gauge-fixing sweeps for a configuration of $SU(2)$ with $a=0.1$
and $\Lambda(x)\neq 0$.\\
In fig. \ref{grf2} it is easy to see that the behaviour of the algorithm when
$\Lambda(x)\neq 0$ is similar to that one with $\Lambda(x)=0$ .\\
Finally we stress that the numerical minimization of the functional 
(\ref{lattf}) is more complicated than that of $F_L(\Omega)$.
In this exploratory study the time per iteration required to minimize $H_L(\Omega)$ is
$5-10$ times the time per iteration required to minimize $F_L(\Omega)$. 

\section{Comments on the lattice Landau gauge-fixing}\label{exappb}
In this section we report some numerical results which show some consequences 
of using different lattice definition of gauge potential in the Landau lattice
gauge-fixing procedure.\\
The standard way of fixing the Landau gauge on the lattice is based on the
minimization of the function $F_L(\Omega)$ defined in the equation (\ref{gribovvecchio})
which is a particular discretization of $F(\Omega)$ with
\be
[A_\mu^2]^{Lat} = -\frac{1}{g^2 a^2}(U_\mu+U^{\dagger}_\mu-2\I)\; .
\ee  
Once $F_L(\Omega)$ has been numerically minimized, the quantity 
\be\label{gribovvecchio2}
\vartheta_F =  \frac{1}{V}\sum_{x} 
\mbox{\rm Tr}[\Delta(x)\Delta^{\dagger}(x)]
\ee
is supposed to vanish and this is the signal that the lattice gauge condition
$\Delta(x)=0$ has been achieved at each site. $\Delta(x)$ is a
discretization of $\partial_\mu A_\mu(x)$ with 
\be
[A_\mu]^{Lat}=\frac{[U\m(x)-U\m^{\dagger}(x)]_{traceless}
}{2iag}\; .
\ee
Some authors showed that, in the lattice Landau Gauge, lattice copies characterized 
by different values of $F_L$ exist. These
solutions of $\Delta^\Omega(x)=0$ are indeed different gauge-related
configurations on the lattice not connected by a colour rotation.
However it is important to note that the definition of the lattice gauge 
potential used in the discretization of $F(\Omega)$ does not
correspond to the definition used in the discretization of the divergence
$\Delta(x)$. This implies that two different lattice definitions of
$A_\mu$ have been used to check the Landau gauge-fixing condition and to 
tag two different solutions.\\  
The authors in ref.\cite{marti1,tasso1} have found two ensembles of
"lattice copies" with $V=16^3\times 32$ and $\beta=6.0$. To analyse
these configurations we used the usual $F_L$ and $\vartheta_F$ variables
defined in equations (\ref{gribovvecchio},\ref{gribovvecchio2})  and another function 
defined as  
\be
\vartheta'_F = \frac{1}{V}\sum_{x} 
\mbox{\rm Tr}[\Delta'(x)\Delta^{' \dagger}(x)]\nonumber
\ee
where 
\begin{eqnarray}
\Delta'(x) & = & \frac{1}{4iag}\sum_{\mu=1}^{4} [(U\m(x)-U\m^{\dagger}(x))
(U\m(x)+U\m^{\dagger}(x))]_{traceless}\nonumber\\
& - & [(U\m(x-\mu)-U\m^{\dagger}(x-\mu))
(U\m(x-\mu)+U\m^{\dagger}(x-\mu))]_{traceless}\nonumber 
\end{eqnarray}
We stress that $\Delta$ and $\Delta'$ are two discretized
definitions of $\partial_\mu A_\mu$ which tend to the same expression as $a\rightarrow 0$. 
For each "lattice copy" we measured
the values of $F_L$, $\vartheta_F$ and $\vartheta'_F$ which we report on 
table \ref{tab2}.
\begin{table}[htb]  
\begin{center}
\begin{tabular}{|c|c|c|c|c|}
\hline
Ensemble & Copy & $F_{L}(\Omega)$ & $\vartheta_F$ & $\vartheta'_F$\\        
\hline
A  & $1$ & $2.583760909874$ & $2.006239333971E-10$ & $0.4114643051772$\\  
   & $2$ & $2.584445214939$ & $3.0326250142058E-10$ & $0.4090001249674$\\  
   & $3$ & $2.584426547617$ & $8.4843522906603E-10$ & $0.4091814689158$\\  
   & $4$ & $2.583991349769$ & $8.2208712589082E-10$ & $0.4106076010266$\\  
   & $5$ & $2.584036530091$ & $2.8552454724294E-10$ & $0.4097852320118$\\    
   & $6$ & $2.58354405348$  & $3.3687550405651E-10$ & $0.4117689556597$\\
\hline
B  & $1$ & $2.581884641098$ & $2.5513621387496E-10$ & $0.4177011836703$\\
   & $2$ & $2.582231848321$ & $7.5376561548352E-10$ & $0.4159162059697$\\
   & $3$ & $2.58231860688$  & $6.1085384335158E-10$ & $0.4159582027108$\\
\hline
\end{tabular}
\end{center}
\caption[]
{Final values of $\vartheta_F$ and $\vartheta'_F$ after gauge-fixing algorithm
which extremizes numerically the functional $F_L$ for two gauge 
fixed ensemble of configuration with $V=16^3\times 32$ and $\beta=6.0$. 
\label{tab2}}
\end{table}
These numerical results show that the finite lattice spacing effects and/or spurious
copy contributions to $\vartheta_F$ and $\vartheta'_F$ are of the order of
$10^{-1}$ while the difference between the values of $F_L$ for two different
copy are of order $10^{-3}$ when $\vartheta_F\leq 10^{-10}$. 
This shows that with this method it is not possible to decide if two 
different solutions of
Landau lattice gauge-fixing condition correspond to different Gribov copies in
the continuum. This procedure is not apt to decide 
if the lattice multiple solution problem of
$\Delta^\Omega(x)=0$ has an analogy with the continuum Gribov problem.\\
Moreover we observe that for a lattice copy there is a big difference between the values 
of the two discretizations $\vartheta_F$ and $\vartheta'_F$ respect to the precision 
required for the minimization. It would be interesting to understand if the difference between
$\vartheta_F$ and $\vartheta'_F$ is only due to the higher order lattice
spacing effects or to spurious copies \cite{forcrand2}.
The problem of higher order contributions and/or spurious solutions 
to $\vartheta_F$ afflicts all
numerical computations of gauge dependent operators. In matching numerical
results obtained on the lattice with the corresponding continuum formulas one must 
carefully evaluate 
the error assigned to the gauge-fixing condition even if $\vartheta\leq
10^{-10}$. Moreover the residual gauge freedom associated to lattice 
copies tagged with the
functional (\ref{gribovvecchio}) should not induce effects 
higher than the sistematic uncertainty due to the higher order contributions
and/or spurious solutions.\\
The numerical results shown in this section indicate also the importance
to improve the gauge-fixing algorithm on the lattice also for the Landau gauge.

\section*{Conclusions}
We have proposed a method which allows the generalization of the Landau 
lattice gauge-fixing procedure to generic covariant gauges. We have shown that
a functional whose stationary points are $\partial_\mu A^\Omega_\mu(x)=\Lambda(x)$ cannot
be obtained as a direct generalization of the Landau lattice gauge-fixing
functional used in literature.  
In the continuum we proposed a functional reaching an extreme 
when $\partial_\mu A^\Omega_\mu(x)=\Lambda(x)$ and we used the simplest
discretization of $H(\Omega)$ to fix numerically a generic covariant gauge on
the lattice. We reported 
preliminary numerical results showing how this procedure works for $SU(2)$ 
and $SU(3)$. Numerical results also show that the contribution of finite 
lattice-spacing effects and/or spurious copies are relevant in the lattice
gauge-fixing procedure and must be carefully evaluate.

\section*{Acknowledgement}
It's a great pleasure to thank M.~Testa for having suggested and supervised
this work. I warmly thank G.~Martinelli, S.~Petrarca, M.~Testa and 
A.~Vladikas for many useful
discussions, suggestions and encouragements throughout this work. 
\section*{Appendix A}
In this appendix we calculate the variation of a $SU(N)$  
matrix
\begin{eqnarray} 
\Omega & = & e^{iw}\\
w & = & \sum_a w^a T^a\nonumber 
\end{eqnarray}
and of the functional 
\be
F(\Omega)=\int d^4x\mbox{\rm Tr}(A^\Omega_{\mu}(x)A^\Omega_{\mu}(x))
\ee
for an infinitesimal variation of $w$ \cite{testa1}.
Following Feynman \cite{feyn}, $\Omega(w+dw)$ is
\begin{eqnarray}
\Omega(w+\delta w) & = & e^{i(w+\delta w)}=e^{i\int^1_0 ds(w_s+\delta
w_s)}\nonumber\\
& = & e^{i\int^1_0 dsw_s} +i\int^1_0 e^{i\int^1_0 dsw_s} \delta w_{s'} 
ds'\nonumber\\ 
& = & \Omega(w)+i\int^1_0 ds'\left[e^{i\int^1_{s'} dsw_s} \delta w_{s'} 
e^{i\int^{s'}_0 dsw_s}\right]\nonumber\\
& = & \Omega(w)+i\int^1_0 ds'\left[e^{i(1-s')w_{s'}} 
\delta w_{s'}e^{is'w}\right]\nonumber\\
& = & \left(\id+i\int^1_0 ds e^{-isw}T^a e^{isw}\delta w^a\right) \Omega(w)\; .
\end{eqnarray}
If we define 
\be
\Sigma^a(s)\equiv e^{-isw}T^a e^{isw}
\ee
and $\sigma^{ab}(s)$ is such a way that
\be
\Sigma^a(s)=\sigma^{ab}(s) T^b
\ee
then it is obvious that
\be
\dot{\Sigma^a}(s)=-i[w,\Sigma^a(s)]
\ee
and then 
\be
\dot{\sigma}^{ab}=w^c f^{cfb}\sigma^{af}.
\ee
If we define $\gamma^{ab}\equiv f^{abc}w^c$ then
\be
\dot{\sigma}=\sigma\gamma \Longrightarrow \sigma=e^{s\gamma}
\ee
and then
\be
\Sigma^a=(e^{\gamma s})^{ab} T^b \Longrightarrow \int^1_0 ds\Sigma^a=\left[
\frac{e^{\gamma}-\id}{\gamma}\right]^{ab} T^b .
\ee
We can conclude that
\begin{eqnarray}\label{ab2}
\Omega(w+\delta w) & = & (\id+i\Phi^{ab}(w)T^b\delta w^a)\Omega(w)\\ 
\Phi^{ab}(w) & \equiv\ & \left[\frac{e^{\gamma}-\id}{\gamma}
\right]^{ab}\nonumber .
\end{eqnarray}
As $\Omega$ is a unitary matrix
\be
\Omega(w+\delta w)\Omega^{\dagger}(w+\delta w)=\I\; ;
\ee
substituting the (\ref{ab2}) in the last equation and considering only linear 
terms in $\delta w$ we obtain
\be
\left(\Phi^{ab}(w) T^b \delta w^a\right)^{\dagger}=\Phi^{ab}(w) T^b \delta 
w^a
\ee
and then
\be\label{ab3}
\Omega^{\dagger}(w+\delta w)=\Omega^{\dagger}(w)(\I+i\Phi^{ab}(w)T^b\delta
w^a)\; .
\ee              
If we remember that 
\begin{eqnarray}\label{ab4}
A^\Omega_{\mu}(x) & = & \Omega(x) A_{\mu}
(x)\Omega^{\dagger}(x)-\frac{i}{g}\Omega(x)\partial_{\mu}
\Omega^{\dagger}(x)\\
\Omega(x) & = & e^{iw(x)}  
\end{eqnarray}
it is easy to verify that 
\be
\delta F(\Omega)=-\frac{2}{g}\int 
d^4x(\partial_{\mu}A^\Omega_{\mu})^a\Phi^{ab}(w)\delta w^b
\ee
and then
\be
\frac{\delta F(\Omega)}{\delta w^b}=-\frac{2}{g}
(\partial_{\mu}A^\Omega_{\mu})^a\Phi^{ab}(w)\; .
\ee

\end{document}